\documentclass[prd,superscriptaddress,altaffilletter,showpacs,nofootinbib]{revtex4}
\usepackage[dvips]{graphicx}
\usepackage{amsmath}
\usepackage{graphicx,epsfig}
\newcommand{\be}{\begin{equation}}
\newcommand{\ee}{\end{equation}}
\newcommand{\bea}{\begin{eqnarray}}
\newcommand{\eea}{\end{eqnarray}}
\newcommand{\der}{\partial}
\newcommand{\vphi}{\varphi}

\begin{document}

\title{Equivalence between Horndeski and beyond Horndeski theories and imperfect fluids}

\author{Israel Quiros}\email{iquiros@fisica.ugto.mx}\affiliation{Dpto. Ingenier\'ia Civil, Divisi\'on de Ingenier\'ia, Universidad de Guanajuato, Gto., M\'exico.}

\author{Ulises Nucamendi}\email{unucamendi@gmail.com}\affiliation{Instituto de F\'isica y Matem\'aticas, Universidad Michoacana de San Nicol\'as de Hidalgo, Edificio C-3, Ciudad Universitaria, CP. 58040 Morelia, Michoac\'an, M\'exico.}\affiliation{Mesoamerican Centre for Theoretical Physics, Universidad Aut\'onoma de Chiapas. Ciudad Universitaria, Carretera Zapata Km. 4, Real del Bosque (Ter\'an), 29040, Tuxtla Guti\'errez, Chiapas, M\'exico.}

\author{Roberto De Arcia}\email{robertodearcia@gmail.com}\affiliation{Dpto. Astronom\'ia, Divisi\'on de Ciencias Exactas, Universidad de Guanajuato, Gto., M\'exico.}

\author{Tame Gonzalez}\email{tamegc72@gmail.com}\affiliation{Dpto. Ingenier\'ia Civil, Divisi\'on de Ingenier\'ia, Universidad de Guanajuato, Gto., M\'exico.}

\author{Francisco Antonio Horta-Rangel}\email{anthort@hotmail.com}\affiliation{Dpto. Ingenier\'ia Civil, Divisi\'on de Ingenier\'ia, Universidad de Guanajuato, Gto., M\'exico.}

\date{\today}

\begin{abstract} In this paper we show that an equivalence between Horndeski and beyond Horndeski theories and general relativity with an effective imperfect fluid can be formally established. The formal equivalence is discussed for several particular cases of interest. Working in the cosmological framework, it is shown that, while the effective stress-energy tensor of viable Horndeski theories is formally equivalent to that of an imperfect fluid with anisotropic stresses and vanishing heat flux vector, the effective stress-energy tensor of beyond Horndeski theories is equivalent to the one of a perfect fluid instead.\end{abstract}


\maketitle


\section{introduction}\label{sec-intro}

Scalar fields have played a very important role in the study of gravitational theories beyond Einstein's general relativity (GR). Among these we may mention the Brans-Dicke (BD) theory \cite{jordan-stt, bd-1961, brans-phd-thesis, goenner-bd-history}, the scalar-tensor theories (STT-s) \cite{fujii-book, faraoni-book, nordvedt-stt, wagoner-stt, bergmann-stt, reasenberg-stt, damour-stt, farese-polarski-prd-2001, anderson-yunes-prd-2017}, the $f(R)$-theory \cite{capozz-fdr, vollick-prd-2003, chiba-fdr, carroll-fdr, sotiriou-cqg-2006, nojiri-odintsov-fdr, cognola-prd-2008, shaw-prd-2008, sotiriou-rmp-2010, defelice-tsujikawa-lrr-2010, nojiri-odintsov-phys-rep-2011, vasilis-phys-rep-2017}, extended theories of gravity (ETG-s) \cite{chiba-jcap-2005, capozziello-grg-rev, capoz-phys-rept-rev, gottlober-cqg-1990, schmidt-cqg-1990, wands-cqg-1994, capoz-etg-grg-2000, nojiri-ijgmp-2007, capoz-etg-prd-2015}, Horndeski \cite{horndeski, nicolis-gal, deffayet-vikman-gal, deffayet-deser-gal, kazuya-silva-gal, fab-4-prl-2012, clifton-phys-rept-rev, deffayet-rev, tsujikawa-lect-not, kazuya-rpp-2016, quiros-ijmpd-rev, kobayashi-rpp-rev} and beyond Horndeski theories \cite{kobayashi-rpp-rev, bhorn-langlois, bhorn-fasiello, bhorn-crisostomi, bhorn(vainsh), mancarella-jcap-2017, bhorn-ostrog, chagoya-tasinato-jhep-2017}. For purpose of comparison with well-understood GR results it is customary to write the field equations of the above mentioned theories  (here we use the units system where $8\pi G_N=1$, with $G_N$ -- the Newton's constant):

\bea G^\text{eff}_NG_{\mu\nu}+{\cal F}_{\mu\nu}\left(R,R_{\sigma\tau}R^{\sigma\tau},R_{\sigma\tau\lambda\kappa}R^{\sigma\tau\lambda\kappa},\nabla^2R,...,\nabla^{2l}R,\phi,\nabla_\mu\phi,\nabla^2\phi,...,\nabla^{2m}\phi\right)=T^\text{mat}_{\mu\nu},\nonumber\eea in the form of Einstein's GR equations where the additional scalar field related and curvature terms are appropriately grouped and organized in the form of an effective stress-energy tensor (SET): 

\bea &&G_{\mu\nu}=\frac{1}{G^\text{eff}_N}\left(T^\text{mat}_{\mu\nu}+T^\text{eff}_{\mu\nu}\right),\nonumber\\
&&T^\text{eff}_{\mu\nu}=-{\cal F}_{\mu\nu}\left(R,R_{\sigma\tau}R^{\sigma\tau},R_{\sigma\tau\lambda\kappa}R^{\sigma\tau\lambda\kappa},\nabla^2R,...,\nabla^{2l}R,\phi,\nabla_\mu\phi,\nabla^2\phi,...,\nabla^{2m}\phi\right).\label{gr-moteq}\eea In the above equations the generic tensor ${\cal F}_{\mu\nu}$ contains the contributions coming from higher-order curvature invariants and/or higher-order derivatives of the curvature scalar $R$ and/or from the scalar field $\phi$ and its higher-order derivatives, while $T^\text{mat}_{\mu\nu}$ accounts for the SET of the matter degrees of freedom: photons, baryons, dark matter, etc. Besides, $R_{\mu\nu}$ is the Ricci tensor, $R^\lambda_{\mu\kappa\nu}$ is the Riemann-Christoffel curvature tensor, $\nabla^2\equiv g^{\mu\nu}\nabla_\mu\nabla_\nu$ and $l$, $m$ are non-vanishing integers. The effective gravitational coupling $G^\text{eff}_N$ in the above equations can be, in principle, a function of the curvature invariants and their higher-order derivatives and of the scalar field and its higher-order derivatives, as well.

The usefulness of Eq. \eqref{gr-moteq} relies on the formal equivalence existing between $T^\text{eff}_{\mu\nu}$ and the SET of perfect and imperfect fluids. The equivalence has been established for scalar-tensor theories \cite{bailyn-prd-1980, madsen-ass-1985, faraoni-prd-2012, semiz-prd-2012, adiez-plb-2013, faraoni-ejpc-2019, madsen-cqg-1988, pimentel-cqg-1989, faraoni-prd-2018} as well as for other modifications of gravity containing higher order derivatives such as the k-essence \cite{adiez-ijmpd-2005, arroja-prd-2010, akhoury-jhep-2009} and its further generalization known as kinetic gravity braiding \cite{pujolas-jcap-2010, pujolas-jhep-2011}, the $f(R)$-theory \cite{capozz-ijgmmp-2018}, the $f(R,G)$ theories ($G$ is the Gauss-Bonnet term) \cite{capozz-ijgmmp-2019} and the ETG-s \cite{capozz-arxiv-2019}. The effective fluid picture has been proved to be useful also within the context of the so called quantum modification of general relativity \cite{novikov}.

It is a well-known fact that when we deal with GR with a minimally coupled self-interacting scalar field $\vphi$, obeying the Einstein's equations of motion (here we omit other matter sources):

\bea G_{\mu\nu}=T^{(\vphi)}_{\mu\nu}=\nabla_\mu\vphi\nabla_\nu\vphi-\frac{1}{2}g_{\mu\nu}(\nabla\vphi)^2-Vg_{\mu\nu},\label{exa-feqs}\eea where $(\nabla\vphi)^2\equiv g^{\mu\nu}\nabla_\mu\phi\nabla_\nu\phi$ and $V=V(\phi)$ is the self-interacting potential, the scalar field's SET $T^{(\vphi)}_{\mu\nu}$ can be written in the equivalent form of a relativistic perfect fluid \cite{ellis, ray-jmp-1972, bailyn-prd-1980, madsen-ass-1985, faraoni-prd-2012, semiz-prd-2012, faraoni-ejpc-2019} after identifying a time-like 4-velocity vector \cite{ellis, madsen-ass-1985, faraoni-prd-2012, semiz-prd-2012}:

\bea u_\mu=\frac{\nabla_\mu\phi}{\sqrt{-(\nabla\phi)^2}},\;-(\nabla\phi)^2\geq 0.\label{4-vel}\eea This time-like vector determines de $3+1$ splitting of the spacetime into a $3$-space seen by co-moving observers and the time direction \cite{ellis}. The metric is written accordingly:

\bea g_{\mu\nu}=h_{\mu\nu}-u_\mu u_\nu=h_{\mu\nu}+\frac{\nabla_\mu\phi\nabla_\nu\phi}{(\nabla\phi)^2},\label{met-split}\eea where $h_{\mu\nu}$ is the metric of the 3-space ($h^\mu_{\;\nu}$ is the projector onto the 3-space orthogonal to the time-direction of co-moving observers):

\bea h^\mu_{\,\lambda}h^\lambda_{\;\nu}=h^\mu_{\;\nu},\;g_{\mu\lambda}h^\lambda_{\,\nu}=h_{\mu\lambda}h^\lambda_{\,\nu}=h_{\mu\nu},\;h_{\mu\nu}u^\nu=0,\;h^\mu_{\;\mu}=3.\label{3-met}\eea Besides, for a given vector $v^\mu$:

\bea h^\lambda_{\;\nu}\nabla_\lambda v_\mu=\nabla_\nu v_\mu+u_\nu u^\lambda\nabla_\lambda v_\mu.\nonumber\eea After the choice \eqref{4-vel} as the 4-velocity of observers co-moving with the scalar field fluid we can rewrite the SET of the scalar field in the form of a perfect fluid SET \cite{ray-jmp-1972, bailyn-prd-1980, madsen-ass-1985, faraoni-prd-2012, semiz-prd-2012, faraoni-ejpc-2019}:

\bea T^{(\vphi)}_{\mu\nu}=-(\nabla\vphi)^2u_\mu u_\nu+\left[-\frac{1}{2}(\nabla\vphi)^2-V\right]g_{\mu\nu}=\left(\rho_\vphi+p_\vphi\right)u_\mu u_\nu+p_\vphi g_{\mu\nu},\label{p-f-set}\eea where we identify:

\bea \rho_\vphi+p_\vphi=-(\nabla\vphi)^2,\;p_\vphi=-\frac{1}{2}(\nabla\vphi)^2-V\;\Rightarrow\;\rho_\vphi=-\frac{1}{2}(\nabla\vphi)^2+V.\label{rho-p-vphi}\eea 

In this effective (perfect fluid) picture $\rho_\vphi$ and $p_\vphi$ represent the energy density and pressure of the fluid. But in general, when non-minimal coupling of the scalar field with the curvature is considered the SET of the scalar field is equivalent to the one of an imperfect fluid as shown in Refs. \cite{madsen-cqg-1988, pimentel-cqg-1989, faraoni-prd-2018} (see the next section). 

It has been shown, also, that theories with higher-order curvature invariants can be written as multi-STT \cite{chiba-jcap-2005}. In particular, the $f(R)$-theory is equivalent to BD theory with vanishing coupling parameter ($\omega=0$) \cite{sotiriou-rmp-2010}. A similar situation arises when theories with higher-order derivatives of the curvature are involved. For instance, theories that are based in the Lagrangian ${\cal L}\propto F\left(R,\nabla^2 R,\nabla^4R,\ldots\nabla^{2k}R\right)$, where $F$ is an arbitrary function of the curvature invariants, are equivalent to multi-STT \cite{wands-cqg-1994}. Take, for instance, the sixth-order gravity given by $F=R+\alpha R\nabla^2 R$, where $\alpha$ is a free constant parameter. This theory can be written as a scalar-tensor theory with Lagrangian:

\bea {\cal L}_\phi=\phi R-\frac{\vphi}{\sqrt{2\alpha}}\left(\phi-1\right)-\frac{1}{2}(\der\vphi)^2.\nonumber\eea This Lagrangian represents Brans-Dicke theory with vanishing coupling parameter $\omega=0$, with BD scalar field $\phi$ and with an additional canonical scalar field $\vphi$ as matter source. 

Given the above mentioned equivalence between theories of gravity containing higher-order curvature terms with (multi) scalar-tensor theories, in what follows we shall omit higher-curvature terms in \eqref{gr-moteq} and we shall focus in theories that include a scalar field and its higher-order derivatives exclusively, i. e., we shall consider theories whose motion equations may be written in the form of the Einstein's GR equation in \eqref{gr-moteq}, with effective SET:

\bea T^\text{eff}_{\mu\nu}=-{\cal F}_{\mu\nu}\left(\phi,\nabla_\mu\phi,\nabla^2\phi,...,\nabla^{2m}\phi\right),\label{case-study}\eea where in ${\cal F}_{\mu\nu}$ we also include possible couplings of the scalar field and/or of its higher-order derivatives with the curvature, as well as self-couplings. This results in generalizations of STT-s known as Horndeski \cite{horndeski, nicolis-gal, deffayet-vikman-gal, deffayet-deser-gal, kazuya-silva-gal, fab-4-prl-2012, clifton-phys-rept-rev, deffayet-rev, tsujikawa-lect-not, kazuya-rpp-2016, kobayashi-rpp-rev, quiros-ijmpd-rev} and beyond Horndeski \cite{bhorn-langlois, bhorn-fasiello, bhorn-crisostomi, bhorn(vainsh), mancarella-jcap-2017, bhorn-ostrog, chagoya-tasinato-jhep-2017, kobayashi-rpp-rev} theories. 

An equivalence between a particular subclass in the Horndeski theories known as kinetic gravity braiding \cite{pujolas-jcap-2010} and general relativity with an effective imperfect fluid has been established in Ref. \cite{pujolas-jhep-2011}. This subclass is given by the following Lagrangian:

\bea {\cal L}_\text{kgb}=K(\phi,X)+G(\phi,X)\nabla^2\phi,\label{pujolas-lag}\eea where $K$ and $G$ are functions of the scalar field $\phi$ and of its kinetic energy: $X\equiv-(\nabla\phi)^2/2$. 

We wonder whether other more general STT-s within the Horndeski and beyond Horndeski classes also admit the above mentioned equivalence with an imperfect fluid. In this paper we are going to show this for a viable Horndeski subclass that is described by the Lagrangian \cite{kobayashi-rpp-rev}:

\bea {\cal L}_\text{vhorn}=G_2(\phi,X)-G_3(\phi,X)\nabla^2\phi+G_4(\phi)\,R,\label{vhorn-lag}\eea where $G_2$ and $G_3$ are functions of $\phi$ and of $X$, while $G_4$ can be a function of $\phi$ only,\footnote{Note that the choice $G_2=K$, $G_3=-G$, $G_4=0$ in \eqref{vhorn-lag} leads to the kinetic gravity braiding model of \eqref{pujolas-lag}.} and also for the viable subclass of beyond Horndeski theories depicted by the Lagrangian:

\bea {\cal L}_\text{vbhorn}=f(\phi,X)R+A(\phi,X)(\nabla X)^2,\label{vbhorn-lag}\eea where $$(\nabla X)^2\equiv\nabla X\cdot\nabla X=\nabla^\lambda\phi\nabla_\mu\nabla_\lambda\phi\nabla^\mu\nabla^\kappa\phi\nabla_\kappa\phi.$$ These subclasses are the only ones that survive the cosmological observational tests, in particular the one related with the nearly simultaneous detection of gravitational waves GW170817 and the $\gamma$-ray burst GRB 170817A \cite{gw-grb}. 

It has to be mentioned that the effective fluid approach to Horndeski theories has been already investigated in Ref. \cite{arjona-prd-2019} within the cosmological setup by means of the cosmological perturbations approach. In the present paper we want to approach the issue from the point of view of relativistic dynamics. Besides, as mentioned, we are going to go further to include the beyond Horndeski theories also.

We have organized the paper in the following way. In the next section we shall apply to the very-well known example of Brans-Dicke theory the procedure we shall use in the paper in order to show the formal equivalence between Horndeski and beyond Horndeski theories and general relativity with an imperfect fluid. In section \ref{sect-3}, for completeness, the basic elements of Horndeski theories are exposed. The formal equivalence between viable Horndeski theories and GR with an imperfect fluid is demonstrated in section \ref{sect-imp-fluid}, while in section \ref{sect-via-bhorn} the mentioned formal equivalence is shown for the viable beyond Horndeski theories. The effective ``imperfect fluid'' picture is explored in the cosmological setting in section \ref{sect-cosmo}. Several important aspects of the explored picture are discussed in section \ref{sect-discuss} where brief conclusions are also given.


\section{the scalar field as an imperfect fluid}\label{sec-2}

As already mentioned, the equations of motion of the scalar-tensor theories -- where the scalar field is non-minimally coupled to the curvature -- can be written as those of GR with an effective imperfect fluid. Here we explain the basis of the formalism in the particular case of the BD theory. The effective imperfect fluid picture for the BD theory has been developed in Refs. \cite{pimentel-cqg-1989, faraoni-prd-2018}. In this case the effective stress-energy tensor is given by:

\bea T_{\mu\nu}^{(\phi)}=\frac{\omega}{\phi^2}\left[\nabla_\mu\phi\nabla_\nu\phi-\frac{1}{2}\,g_{\mu\nu}(\nabla\phi)^2\right]-\frac{V}{2\phi}g_{\mu\nu}+\frac{1}{\phi}\left(\nabla_\mu\nabla_\nu\phi-g_{\mu\nu}\nabla^2\phi\right),\label{set-phi}\eea where $\omega$ is the BD coupling parameter and $V$ is the self-interacting potential for the scalar field.

We want to show that the above effective SET can be written in the form of the stress-energy tensor of an imperfect fluid:

\bea T_{\mu\nu}^\text{(if)}=(\rho+p)u_\mu u_\nu+pg_{\mu\nu}+2q_{(\mu}u_{\nu)}+\pi_{\mu\nu},\label{set-imp-f}\eea where $\rho$, $p$ are the energy density and pressure of the fluid, $q_\mu$ is the heat flux vector and $\pi_{\mu\nu}$ is the anisotropic SET: 

\bea \pi_{\mu\nu}=\Pi_{\mu\nu}-\frac{1}{3}\,\Pi h_{\mu\nu},\;\pi=\pi^\mu_{\;\mu}=0,\;\pi_{\mu\nu}u^\nu=0,\label{set-anisot}\eea with

\bea \Pi_{\mu\nu}=T_{\lambda\kappa}h^\lambda_{\;\mu}h^\kappa_{\;\nu}=ph_{\mu\nu}+\pi_{\mu\nu},\;\Pi=\Pi^\mu_{\;\mu}=3p,\;\Pi_{\mu\nu}u^\nu=0.\label{set-Pi}\eea

By comparing equations \eqref{set-phi} and \eqref{set-imp-f} one gets that

\bea (\rho+p)u_\mu u_\nu=\frac{\omega}{\phi^2}\nabla_\mu\phi\nabla_\nu\phi\,\Rightarrow\,u_\mu u_\nu=-\frac{\nabla_\mu\phi\nabla_\nu\phi}{(\nabla\phi)^2},\nonumber\eea from where the time-like 4-velocity vector of the fluid $u^\mu$ is defined as in \eqref{4-vel}. Following the approach explained in the introduction, the energy density and pressure of the fluid as well as the heat flux vector $q_\mu$, are defined as it follows:

\bea \rho=T_{\mu\nu}u^\mu u^\nu,\;p=\frac{1}{3}\,\Pi,\;q_\mu=-T_{\lambda\kappa}u^\lambda h^\kappa_{\;\mu}.\label{rho-p-q}\eea Other kinematic quantities of the fluid are the following:

\bea &&\dot u_\mu=u^\nu\nabla_\nu u_\mu,\nonumber\\
&&\theta=\nabla_\mu u^\mu,\nonumber\\
&&\sigma_{\mu\nu}=\nabla_{(\mu}u_{\nu)}+\dot u_{(\mu}u_{\nu)}-\frac{1}{3}\theta h_{\mu\nu},\nonumber\\
&&\omega_{\mu\nu}=\nabla_{[\mu}u_{\nu]}+\dot u_{[\mu}u_{\nu]},\label{kin-q}\eea where $\dot u_\mu$ is the acceleration of the fluid, $\theta$ is the expansion, $\sigma_{\mu\nu}$ is the shear tensor of the fluid, while $\omega_{\mu\nu}$ accounts for the vorticity tensor. Under the choice \eqref{4-vel}, since the 4-velocity is the gradient of a scalar, the vorticity tensor $\omega_{\mu\nu}$ vanishes identically. This is true for any scalar-tensor theories and their higher-derivative modifications: Horndeski and beyond Horndeski theories.  

When we take into account the definition of the stress-energy tensor of the BD scalar field \eqref{set-phi}, the tensor \eqref{set-Pi} is given by \cite{faraoni-prd-2018}:

\bea &&\Pi^{(\phi)}_{\mu\nu}=-\left[\frac{\omega}{2\phi^2}(\nabla\phi)^2+\frac{V}{2\phi}+\frac{2}{3}\frac{\nabla^2\phi}{\phi}+\frac{\nabla^\kappa\phi\nabla^\lambda\phi\nabla_\lambda\nabla_\kappa\phi}{3\phi(\nabla\phi)^2}\right]h_{\mu\nu}+\frac{1}{\phi}\left[\nabla_\mu\nabla_\nu-\frac{1}{3}h_{\mu\nu}\nabla^2\right]\phi\nonumber\\
&&\;\;\;\;\;\;\;\;\;\;\;\;-\frac{\nabla^\lambda\phi}{\phi(\nabla\phi)^2}\left[\nabla_\lambda\nabla_\mu\phi\nabla_\nu\phi+\nabla_\lambda\nabla_\nu\phi\nabla_\mu\phi-\frac{1}{3}h_{\mu\nu}\nabla^\kappa\phi\nabla_\lambda\nabla_\kappa\phi-\frac{\nabla_\mu\phi\nabla_\nu\phi\nabla^\kappa\phi\nabla_\lambda\nabla_\kappa\phi}{(\nabla\phi)^2}\right].\label{set-Pi-phi}\eea If we compare this latter equation with \eqref{set-Pi}, for the anisotropic SET we obtain:

\bea &&\pi^{(\phi)}_{\mu\nu}=-\frac{\nabla^\lambda\phi}{\phi(\nabla\phi)^2}\left[\nabla_\lambda\nabla_\mu\phi\nabla_\nu\phi+\nabla_\lambda\nabla_\nu\phi\nabla_\mu\phi-\frac{1}{3}h_{\mu\nu}\nabla^\kappa\phi\nabla_\lambda\nabla_\kappa\phi-\frac{\nabla_\mu\phi\nabla_\nu\phi\nabla^\kappa\phi\nabla_\lambda\nabla_\kappa\phi}{(\nabla\phi)^2}\right]\nonumber\\
&&\;\;\;\;\;\;\;\;\;\;\;\;\;\;\;\;\;\;\;\;\;\;\;\;\;\;\;\;\;\;\;\;\;\;\;\;\;\;\;\;\;\;\;\;\;\;\;\;\;\;\;\;\;\;\;\;\;\;\;\;\;\;\;\;\;\;\;\;\;\;\;\;\;\;\;\;\;\;\;\;\;\;\;\;\;\;\;\;\;\;\;\;\;\;\;\;\;\;\;\;\;\;\;\;\;\;\;+\frac{1}{\phi}\left[\nabla_\mu\nabla_\nu-\frac{1}{3}h_{\mu\nu}\nabla^2\right]\phi,\label{set-pi-phi}\eea while for the pressure of the fluid:

\bea p_\phi=-\left[\frac{\omega}{2\phi^2}(\nabla\phi)^2+\frac{V}{2\phi}+\frac{2}{3}\frac{\nabla^2\phi}{\phi}+\frac{\nabla^\kappa\phi\nabla^\lambda\phi\nabla_\lambda\nabla_\kappa\phi}{3\phi(\nabla\phi)^2}\right].\label{p-phi}\eea For other relevant quantities appearing in \eqref{set-imp-f} we get:

\bea \rho_\phi=-\left[\frac{\omega}{2\phi^2}(\nabla\phi)^2-\frac{V}{2\phi}-\frac{\nabla^2\phi}{\phi}+\frac{\nabla^\kappa\phi\nabla^\lambda\phi\nabla_\lambda\nabla_\kappa\phi}{\phi(\nabla\phi)^2}\right],\label{rho-phi}\eea for the energy density of the BD field, while for the heat flux vector \cite{faraoni-prd-2018}:

\bea q^{(\phi)}_\mu=-\frac{\nabla^\lambda\phi}{\phi\sqrt{-(\nabla\phi)^2}}\left[\nabla_\lambda\nabla_\mu\phi-\frac{\nabla_\mu\phi\nabla^\kappa\phi\nabla_\kappa\nabla_\lambda\phi}{(\nabla\phi)^2}\right]=-\frac{\sqrt{-(\nabla\phi)^2}}{\phi}\,\dot u_\mu,\label{q-flux}\eea where we have taken into account that for the choice \eqref{4-vel} the acceleration of the fluid is given by:

\bea \dot u_\mu=u^\lambda\nabla_\lambda u_\mu=-\frac{\nabla^\lambda\phi}{(\nabla\phi)^2}\left[\nabla_\lambda\nabla_\mu\phi-\frac{\nabla_\mu\phi\nabla^\kappa\phi\nabla_\kappa\nabla_\lambda\phi}{(\nabla\phi)^2}\right].\label{accel-phi}\eea

Below we shall apply this formalism to the Horndeski and beyond Horndeski theories in order to show that a similar effective imperfect fluid picture is possible.


\section{horndeski theories}\label{sect-3}

According to \cite{deffayet_prd_2011}, the most general 4-dimensional scalar-tensor theories having second-order motion equations are described by the linear combinations of the following Lagrangians:

\bea &&{\cal L}_2=K,\;{\cal L}_3 =-G_3(\nabla^2\phi),\;{\cal L}_4=G_4 R+G_{4,X}\left[(\nabla^2\phi)^2-(\nabla_\mu\nabla_\nu\phi)^2\right],\nonumber\\
&&{\cal L}_5=G_5 G_{\mu\nu}\nabla^{\mu}\nabla^\nu\phi-\frac{1}{6}G_{5,X}\left[(\nabla^2\phi)^3-3\nabla^2\phi(\nabla_\mu\nabla_\nu\phi)^2+2(\nabla_\mu\nabla_\nu\phi)^3\right],\label{horn-lags}\eea where $K=K(\phi,X)$ and $G_i=G_i(\phi,X)$ ($i=3,4,5$), are functions of the scalar field $\phi$ and of its kinetic energy density $X$, while $G_{i,\phi}$ and $G_{i,X}$, represent the derivatives of the functions $G_i$ with respect to $\phi$ and $X$, respectively. In the Lagrangian ${\cal L}_5$ above, for compactness of writing, we have adopted the same definitions used in Ref. \cite{kobayashi-rpp-rev}:
 
\bea (\nabla_\mu\nabla_\nu\phi)^2&:=\nabla_\mu\nabla_\nu\phi\nabla^\mu\nabla^\nu\phi,\;(\nabla_\mu\nabla_\nu\phi)^3&:=\nabla^\mu\nabla_\alpha\phi\nabla^\alpha\nabla_\beta\phi\nabla^\beta\nabla_\mu\phi.\label{def}\eea Note that in \eqref{horn-lags} we have slightly modified the notation with respect to \eqref{vhorn-lag}, since we have replaced $K\rightarrow G_2$. We have done this in order to meet the notation most frequently found in the bibliography.

The general action for the Horndeski theories can be written as:

\bea S_\text{Horn}=\int d^4x\sqrt{|g|}\left({\cal L}_2+{\cal L}_3+{\cal L}_4+{\cal L}_5+{\cal L}_\text{mat}\right),\label{horn-action}\eea where the ${\cal L}_i$ ($i=2,3,4$) are given by \eqref{horn-lags} and ${\cal L}_m$ stands for the Lagrangian of the matter degrees of freedom. The motion equations that can be derived from the above action read:

\bea G_{\mu\nu}=\frac{1}{2G_4}\,T^\text{mat}_{\mu\nu}+\sum_i T^{(i)}_{\mu\nu},\;i=2,3,4,\label{horn-moteq}\eea where $$\frac{\delta\left(\sqrt{|g|}{\cal L}_\text{mat}\right)}{\sqrt{|g|}\delta g^{\mu\nu}}=-\frac{1}{2}\,T^\text{mat}_{\mu\nu},$$ and we have considered the following definitions of the effective stress-energy tensors related with the ${\cal L}_i$-s in \eqref{horn-lags}:

\bea &&T^{(2)}_{\mu\nu}=\frac{1}{2G_4}\left(K_{,X}\nabla_\mu\phi\nabla_\nu\phi+K g_{\mu\nu}\right),\nonumber\\
&&T^{(3)}_{\mu\nu}=\frac{1}{2G_4}\left\{-\left(2G_{3,\phi}+G_{3,X}\nabla^2\phi\right)\nabla_\mu\phi\nabla_\nu\phi-2G_{3,X}\nabla_{(\mu}\phi\nabla_{\nu)} X+g_{\mu\nu}\left[G_{3,\phi}(\nabla\phi)^2+G_{3,X}(\nabla\phi\cdot\nabla X)\right]\right\},\nonumber\\
&&T^{(4)}_{\mu\nu}=\frac{G_{4,\phi}}{G_4}\left(\nabla_\mu\nabla_\nu\phi-g_{\mu\nu}\nabla^2\phi\right)+\frac{G_{4,\phi\phi}}{G_4}\left[\nabla_\mu\phi\nabla_\nu\phi-g_{\mu\nu}(\nabla\phi)^2\right],\label{set-234}\eea where $(\nabla\phi\cdot\nabla X)\equiv g^{\mu\nu}\nabla_\mu\phi\nabla_\nu X$, with $\nabla_\mu X\equiv-\nabla^\lambda\phi(\nabla_\mu\nabla_\lambda\phi)$. Notice that in the definition of the SET-s $T^{(i)}_{\mu\nu}$ above we have already included the contribution coming from the effective gravitational coupling $G_4$. Besides, since the latter is a function of the scalar field only: $G_4=G_4(\phi)$, the resulting theory is in the viable subclass of Horndeski theories \eqref{vhorn-lag} mentioned in the introduction. For the same reason we have not considered the contribution coming from the Lagrangian ${\cal L}_5$ in \eqref{horn-action}.


\section{equivalence between viable Horndeski theories and imperfect fluids}\label{sect-imp-fluid}

Here, as before, we consider the time-like 4-velocity vector defined as in Eq. \eqref{4-vel}: $u^\mu=\nabla^\mu\phi/\sqrt{2X}$, with non-negative $X\geq 0$, in order to determine the $3+1$ splitting of the spacetime (recall that $X\equiv-(\nabla\phi)^2/2$ is the kinetic energy of the scalar field). Given the adopted definition of the time-like 4-velocity \eqref{4-vel}, we can rewrite the kinematic quantities in \eqref{kin-q} in terms of our notation as it follows. For the expansion we have:

\bea \theta=\frac{1}{\sqrt{2X}}\left[\nabla^2\phi-\frac{(\nabla\phi\cdot\nabla X)}{2X}\right],\label{exp-x}\eea while for the components of the shear and vorticity tensors:

\bea &&\sigma_{\mu\nu}=\frac{1}{\sqrt{2X}}\left\{\nabla_\mu\nabla_\nu\phi-\frac{(\nabla\phi\cdot\nabla X)}{4X^2}\nabla_\mu\phi\nabla_\nu\phi-\frac{\nabla_{(\nu}\phi\nabla_{\mu)}X}{X}-\frac{\theta}{3}h_{\mu\nu}\right\},\nonumber\\
&&\omega_{\mu\nu}=\frac{2}{(2X)^{3/2}}\nabla_{[\mu}\phi\nabla_{\nu]}X,\label{shear-vort-x}\eea respectively. Above we taken into account that

\bea \nabla_\mu u_\nu=\frac{1}{\sqrt{2X}}\left(\nabla_\mu\nabla_\nu\phi-\frac{\nabla_\nu\phi\nabla_\mu X}{2X}\right).\label{n-u}\eea so that the acceleration can be written as:

\bea \dot u_\mu=u^\lambda\nabla_\lambda u_\mu=-\frac{1}{2X}\left[\nabla_\mu X+\frac{(\nabla\phi\cdot\nabla X)}{2X}\nabla_\mu\phi\right].\label{accel-x}\eea 

In order to show the equivalence between the viable Horndeski theories \eqref{horn-moteq}, \eqref{set-234} and imperfect fluids we shall consider each of the effective SET-s in \eqref{set-234} separately. Let us start with $T^{(2)}_{\mu\nu}$. This effective tensor corresponds to the so called k-essence models. Although the equivalence between these models and a perfect fluid has been already demonstrated \cite{adiez-ijmpd-2005, arroja-prd-2010}, here we write the basic equations in terms of our notation. Following the procedure exposed in section \ref{sec-2} we obtain the following results:

\bea &&\Pi^{(2)}_{\mu\nu}=T^{(2)}_{\lambda\kappa}h^\lambda_{\;\mu}h^\kappa_{\;\nu}=\frac{K}{2G_4}\,h_{\mu\nu},\;\pi^{(2)}_{\mu\nu}=0,\;q^{(2)}_\mu=-T^{(2)}_{\lambda\kappa}u^\lambda h^\kappa_{\;\mu}=0,\nonumber\\
&&p_{(2)}=\frac{1}{3}\Pi^{(2)}=\frac{K}{2G_4},\;\rho_{(2)}=T^{(2)}_{\lambda\kappa}u^\lambda u^\kappa=\frac{1}{2G_4}\left(2XK_{,X}-K\right).\label{k-essence}\eea 

In what regards to the piece $T^{(3)}_{\mu\nu}$, the calculations are a bit more complicated. Let us to start by computing the tensor:

\bea \Pi^{(3)}_{\mu\nu}=T^{(3)}_{\lambda\kappa}h^\lambda_{\;\mu}h^\kappa_{\;\nu}=-\frac{1}{G_4}\left[G_{3,\phi}X-\frac{G_{3,X}}{2}(\nabla\phi\cdot\nabla X)\right]h_{\mu\nu},\label{Pi3}\eea so that the effective pressure $p_{(3)}$ is given by:

\bea p_{(3)}=\frac{1}{3}\Pi^{(3)}=-\frac{1}{G_4}\left[G_{3,\phi}X-\frac{G_{3,X}}{2}(\nabla\phi\cdot\nabla X)\right],\label{p3}\eea meanwhile the calculation of effective energy density $\rho_{(3)}$ gives:

\bea \rho_{(3)}=T^{(3)}_{\lambda\kappa}u^\lambda u^\kappa=-\frac{1}{G_4}\left[G_{3,\phi}X-\frac{G_{3,X}}{2}(\nabla\phi\cdot\nabla X)+G_{3,X} X(\nabla^2\phi)\right].\label{rho3}\eea It can be shown that $\pi^{(3)}_{\mu\nu}=0$, so that the effective fluid does not have anisotropic stresses. However, there is a non-vanishing heat flux given by:

\bea q^{(3)}_\mu=-\frac{G_{3,X}}{2\sqrt{2X}G_4}\left[2X\nabla_\mu X+\nabla_\mu\phi(\nabla\phi\cdot\nabla X)\right].\label{q3}\eea

The effective SET tensor $T^{(4)}_{\mu\nu}$ in Eq. \eqref{set-234} can be written in the alternative way:

\bea T^{(4)}_{\mu\nu}=\frac{G_{4,\phi}}{G_4}\left(\nabla_\mu\nabla_\nu\phi-g_{\mu\nu}\nabla^2\phi\right)+\frac{G_{4,\phi\phi}}{G_4}2Xh_{\mu\nu}.\label{tmn4}\eea Hence

\bea \Pi^{(4)}_{\mu\nu}=T^{(4)}_{\lambda\kappa}h^\lambda_{\;\mu}h^\kappa_{\;\nu}=\left(\frac{G_{4,\phi\phi}}{G_4}2X-\frac{G_{4,\phi}}{G_4}\nabla^2\phi\right)h_{\mu\nu}+\frac{G_{4,\phi}}{G_4}\left[\nabla_\mu\nabla_\nu\phi-\frac{1}{X}\nabla_{(\mu}\phi\nabla_{\nu)}X-\frac{(\nabla\phi\cdot\nabla X)}{4X^2}\nabla_\mu\phi\nabla_\nu\phi\right].\label{Pi4}\eea Since the resulting effective pressure $p_{(4)}=\Pi_{(4)}/3$, we get:

\bea p_{(4)}=-\frac{G_{4,\phi}}{G_4}\left[\frac{(\nabla\phi\cdot\nabla X)}{2X}+\frac{2}{3}\nabla^2\phi\right]+\frac{G_{4,\phi\phi}}{G_4}2X.\label{p4}\eea This entails that, since $\pi^{(4)}_{\mu\nu}=\Pi^{(4)}_{\mu\nu}-p_{(4)}h_{\mu\nu}$, for the tensor of anisotropic stresses we obtain the following expression:

\bea \pi^{(4)}_{\mu\nu}=\frac{G_{4,\phi}}{G_4}\left\{\left[\frac{(\nabla\phi\cdot\nabla X)}{2X}-\frac{1}{3}\nabla^2\phi\right]h_{\mu\nu}+\nabla_\mu\nabla_\nu\phi-\frac{1}{X}\nabla_{(\mu}\phi\nabla_{\nu)}X-\frac{(\nabla\phi\cdot\nabla X)}{4X^2}\nabla_\mu\phi\nabla_\nu\phi\right\}.\label{pi4}\eea 

The effective energy density is given by:

\bea \rho_{(4)}=T^{(4)}_{\lambda\kappa}u^\lambda u^\kappa=-\frac{G_{4,\phi}}{G_4}\left[\frac{(\nabla\phi\cdot\nabla X)}{2X}-\nabla^2\phi\right],\label{rho4}\eea while the heat flux vector:

\bea q^{(4)}_\mu=-T^{(4)}_{\lambda\kappa}u^\lambda h^\kappa_{\;\mu}=\frac{G_{4,\phi}}{\sqrt{2X}G_4}\left[\nabla_\mu X+\frac{(\nabla\phi\cdot\nabla X)}{2X}\nabla_\mu\phi\right].\label{q4}\eea

In consequence the viable Horndeski theory \eqref{horn-moteq} is equivalent to GR with the effective stress-energy tensor of an imperfect fluid:

\bea T_{\mu\nu}^\text{eff}=(\rho_\text{eff}+p_\text{eff})u_\mu u_\nu+p_\text{eff}g_{\mu\nu}+2q^\text{eff}_{(\mu}u_{\nu)}+\pi^\text{eff}_{\mu\nu},\label{eff-set}\eea where

\bea \rho_\text{eff}=\sum_i\rho_{(i)},\;p_\text{eff}=\sum_i p_{(i)},\;q^\text{eff}_\mu=\sum_i q^{(i)}_\mu,\;\pi^\text{eff}_{\mu\nu}=\sum_i\pi^{(i)}_{\mu\nu},\;i=2,3,4.\label{eff-qs}\eea 

We have, in particular, that for the effective heat flux vector (for the effective pressure and energy density see equations \eqref{p-eff-vhorn} and \eqref{rho-eff-vhorn} in section \ref{sect-discuss}):

\bea q^\text{eff}_\mu=\frac{G_{4,\phi}-XG_{3,X}}{\sqrt{2X}G_4}\left[\nabla_\mu X+\frac{(\nabla\phi\cdot\nabla X)}{2X}\nabla_\mu\phi\right],\label{q-eff-vhorn}\eea while the effective anisotropic stress coincides with $\pi^{(4)}_{\mu\nu}$ in \eqref{pi4}: 

\bea \pi^\text{eff}_{\mu\nu}=\frac{G_{4,\phi}}{G_4}\left\{\left[\frac{(\nabla\phi\cdot\nabla X)}{2X}-\frac{1}{3}\nabla^2\phi\right]h_{\mu\nu}+\nabla_\mu\nabla_\nu\phi-\frac{1}{X}\nabla_{(\mu}\phi\nabla_{\nu)}X-\frac{(\nabla\phi\cdot\nabla X)}{4X^2}\nabla_\mu\phi\nabla_\nu\phi\right\}.\label{pi-eff-vhorn}\eea Non-vanishing either $q^\text{eff}_\mu$ or $\pi^\text{eff}_{\mu\nu}$ or both, is what distinguishes an imperfect effective fluid from a perfect one. This means that, while the heat flux is due both to the non-minimal coupling $G_4=G_4(\phi)$ and to the higher-derivative coupling $G_3=G_3(\phi, X)$, the anisotropic stresses are the consequence of the non-minimal coupling only. In particular, for constant $G_4=1/2$ (minimal coupling), the anisotropic stresses vanish.


\subsection{Particular cases}\label{subsect-cases}

Les us check several particular cases in the Horndeski class of theories \cite{quiros-ijmpd-rev, tsujikawa_lect_not}:

\begin{enumerate}

\item{\it General relativity with a minimally coupled scalar field.} This is given by the following choice of the relevant functions in \eqref{horn-lags}: $G_4=1/2,\;G_3=G_5=0,$
 
\bea S=\int d^4x\sqrt{|g|}\left[\frac{1}{2}\,R+K(\phi,X)+{\cal L}_m\right].\nonumber\eea This choice comprises quintessence; $K(\phi,X)=X-V$, and k-essence, for instance, $K(\phi,X)=f(\phi)g(X)$, where $f$ and $g$ are arbitrary functions of their arguments. The most important kinematic quantities for this case are given in Eq. \eqref{k-essence}:

\bea \Pi^\text{eff}_{\mu\nu}=Kh_{\mu\nu},\;\pi^\text{eff}_{\mu\nu}=0,\;q^\text{eff}_\mu=0,\;p_\text{eff}=K,\;\rho_\text{eff}=2XK_{,X}-K.\nonumber\eea

\item{\it Brans-Dicke theory.} The following choice corresponds to the BD theory \cite{bd-1961} (here $\omega$ is the BD coupling parameter): $K(\phi,X)=2\omega X/\phi-V(\phi),\;G_3=G_5=0,\;G_4=\phi,$

\bea S=\int d^4x\sqrt{|g|}\left[\phi R+2\omega X/\phi-V\right].\label{bd-action}\eea Although the kinematic quantities for this case have been already computed in section \ref{sec-2}, here we rewrite them in terms of the present notation. In this case, the quantities \eqref{eff-qs} in the effective stress energy tensor \eqref{eff-set} read:

\bea &&\rho_\text{eff}=\frac{\omega X}{\phi^2}+\frac{V}{2\phi}+\frac{\nabla^2\phi}{\phi}-\frac{(\nabla\phi\cdot\nabla X)}{2\phi X},\;p_\text{eff}=\frac{\omega X}{\phi^2}-\frac{V}{2\phi}-\frac{2\nabla^2\phi}{3\phi}-\frac{(\nabla\phi\cdot\nabla X)}{2\phi X},\nonumber\\
&&\pi^\text{eff}_{\mu\nu}=\left[\frac{(\nabla\phi\cdot\nabla X)}{2\phi X}-\frac{\nabla^2\phi}{3\phi}\right]h_{\mu\nu}+\frac{\nabla_\mu\nabla_\nu\phi}{\phi}-\frac{\nabla_{(\mu}\phi\nabla_{\nu)}X}{\phi X}-\frac{(\nabla\phi\cdot\nabla X)}{4\phi X^2}\nabla_\mu\phi\nabla_\nu\phi,\nonumber\\
&&q^\text{eff}_\mu=\frac{1}{\phi\sqrt{2X}}\left[\nabla_\mu X+\frac{(\nabla\phi\cdot\nabla X)}{2X}\nabla_\mu\phi\right].\label{bd-q}\eea Notice that the above expressions coincide with the corresponding ones in \eqref{set-pi-phi}-\eqref{q-flux}.

\item{\it Non-minimal coupling (NMC) theory.} This is described by the functions: $K=\omega(\phi)X-V(\phi),\;G_4=(1-\xi\phi^2)/2,\;G_3=G_5=0,$

\bea S=\int dx^4\sqrt{|g|}\left[\frac{1-\xi\phi^2}{2}R+\omega(\phi)X-V(\phi)\right].\nonumber\eea The main kinematic quantities in \eqref{eff-set} are:

\bea &&\rho_\text{eff}=\frac{1}{1-\xi\phi^2}\left\{\omega X+V-2\xi\phi\nabla^2\phi+2\xi\phi\frac{(\nabla\phi\cdot\nabla X)}{2X}\right\},\nonumber\\
&&p_\text{eff}=\frac{1}{1-\xi\phi^2}\left\{\omega X-V+\frac{4\xi\phi}{3}\nabla^2\phi+2\xi\phi\frac{(\nabla\phi\cdot\nabla X)}{2X}-4\xi X\right\},\nonumber\\
&&\pi^\text{eff}_{\mu\nu}=-\frac{2\xi\phi}{1-\xi\phi^2}\left\{\left[\frac{(\nabla\phi\cdot\nabla X)}{2X}-\frac{1}{3}\nabla^2\phi\right]h_{\mu\nu}+\frac{\nabla_\mu\nabla_\nu\phi}{\phi}-\frac{\nabla_{(\mu}\phi\nabla_{\nu)}X}{\phi X}-\frac{(\nabla\phi\cdot\nabla X)}{4\phi X^2}\nabla_\mu\phi\nabla_\nu\phi\right\},\nonumber\\
&&q^\text{eff}_\mu=-\frac{2\xi\phi}{(1-\xi\phi^2)\sqrt{2X}}\left[\nabla_\mu X+\frac{(\nabla\phi\cdot\nabla X)}{2X}\nabla_\mu\phi\right].\label{nmc-q}\eea

\item{\it Cubic galileon.} For this particular case in the functions in \eqref{horn-lags} one sets: $K=2\omega X/\phi-2\Lambda\phi,$ $G_3=-2f(\phi)X,$ $G_4=\phi,\;G_5=0,$ and the resulting Jordan frame (JF) action reads \cite{kazuya_gal}: 

\bea S=\int d^4x\sqrt{|g|}\left[\phi R+2\omega X/\phi-2\Lambda\phi-2Xf(\phi)\nabla^2\phi\right].\nonumber\eea We obtain the following expressions for the fundamental kinematic quantities in \eqref{eff-set}, \eqref{eff-qs}:

\bea &&\rho_\text{eff}=\frac{\omega X}{\phi^2}+\Lambda+2\frac{f_{,\phi}}{\phi}X^2+\left[1+2f(\phi)X\right]\left[\frac{\nabla^2\phi}{\phi}-\frac{(\nabla\phi\cdot\nabla X)}{2\phi X}\right],\nonumber\\
&&p_\text{eff}=\frac{\omega X}{\phi^2}+\Lambda+2\frac{f_{,\phi}}{\phi}X^2-\frac{2\nabla^2\phi}{3\phi}-\left[1+2f(\phi)X\right]\frac{(\nabla\phi\cdot\nabla X)}{2\phi X},\nonumber\\
&&\pi^\text{eff}_{\mu\nu}=\left[\frac{(\nabla\phi\cdot\nabla X)}{2\phi X}-\frac{\nabla^2\phi}{3\phi}\right]h_{\mu\nu}+\frac{\nabla_\mu\nabla_\nu\phi}{\phi}-\frac{\nabla_{(\mu}\phi\nabla_{\nu)}X}{\phi X}-\frac{(\nabla\phi\cdot\nabla X)}{4\phi X^2}\nabla_\mu\phi\nabla_\nu\phi,\nonumber\\
&&q^\text{eff}_\mu=\frac{1+2f(\phi)X}{\phi\sqrt{2X}}\left[\nabla_\mu X+\frac{(\nabla\phi\cdot\nabla X)}{2X}\nabla_\mu\phi\right].\label{cubic-jf-q}\eea
 
\end{enumerate} 

The above examples belong in the viable Horndeski theories \eqref{vhorn-lag}, where by viable we mean that the speed of propagation of the tensor perturbations coincides with the speed of light: $c^2_\text{gw}=1$. However, there are other interesting cases that do not belong in \eqref{vhorn-lag} and, hence, are not considered in the present paper. One very interesting example is given by the choice: $K=X-V,\;G_3=0,\;G_4=1/2,\;G_5=-\alpha\phi/2.$ The corresponding theory is known as kinetic coupling to the Einstein's tensor. The resulting action reads:

\bea S=\frac{1}{2}\int d^4x\sqrt{|g|}\left[R+2(X-V)+\alpha G_{\mu\nu}\der^\mu\phi\der^\nu\phi\right].\nonumber\eea Another example that does not belong in the viable Horndeski theories \eqref{vhorn-lag} is the so called covariant galileon \cite{cov-gal}. This theory corresponds to the following choice in \eqref{horn-lags}: $$K=-2c_2,\;G_3=-2c_3 X,\;G_4=\frac{1}{2}-4c_4 X^2,\;G_5=-4c_5 X^2,$$ where the $c_i$-s are constants.


\section{Viable beyond Horndeski theories as imperfect fluids}\label{sect-via-bhorn}

Here we consider the viable beyond Horndeski theories \cite{bhorn} (also known as degenerate higher-order scalar-tensor theories) with Lagrangian \eqref{vbhorn-lag}. The following effective Einstein's equation can be derived from the latter Lagrangian: $G_{\mu\nu}=T^\text{vbhorn}_{\mu \nu},$ where the effective SET for the viable beyond Horndeski fluid is given by the following expression:

\bea &&T^\text{vbhorn}_{\mu \nu}=\frac{1}{2f}\left[f_{,X}R-A_{,X}(\nabla X)^2-2A_{,\phi}(\nabla\phi\cdot\nabla X)-2A\nabla^2 X\right]\nabla_\mu\phi\nabla_\nu\phi\nonumber\\
&&\;\;\;\;\;\;\;\;\;\;\;\;+\frac{f_{,\phi\phi}}{f}\left[\nabla_\mu\phi\nabla_\nu\phi+2Xg_{\mu\nu}\right]+\frac{f_{,\phi}}{f}\left(\nabla_\mu\nabla_\nu\phi-g_{\mu\nu}\nabla^2\phi\right)\nonumber\\
&&\;\;\;\;\;\;\;\;\;\;\;\;+2\frac{f_{,\phi X}}{f}\left[\nabla_{(\mu}\phi\nabla_{\nu)}X-g_{\mu\nu}(\nabla\phi\cdot\nabla X)\right]+\frac{(f_{,XX}-A)}{f}\nabla_\mu X\nabla_\nu X\nonumber\\
&&\;\;\;\;\;\;\;\;\;\;\;\;-\frac{(2f_{,XX}-A)}{2f}g_{\mu\nu}(\nabla X)^2+\frac{f_{,X}}{f}\left(\nabla_\mu\nabla_\nu X-g_{\mu\nu}\nabla^2X\right),\label{vbhorn-set}\eea where $$A=\frac{3f_{,X}^2}{2f};\;A_{,X}=3\left(\frac{f_{,X}}{f}\right)f_{,XX}-\frac{3}{2}\left(\frac{f_{,X}}{f}\right)^2 f_{,X}.$$ The above stress-energy tensor \eqref{vbhorn-set} can be written in the form of an effective SET for an imperfect fluid \eqref{eff-set} with effective energy density:

\bea &&\rho_\text{eff}=\frac{f_{,X}}{f}XR+\frac{f_{,\phi}}{f}\nabla^2\phi+\frac{2f_{,XX}-A-2XA_{,X}}{2f}(\nabla X)^2+\frac{f_{,X}-2XA}{f}\nabla^2X\nonumber\\
&&\;\;\;\;\;\;\;-\frac{f_{,\phi}+4X^2A_{,\phi}}{2Xf}(\nabla\phi\cdot\nabla X)+\frac{f_{,XX}-A}{2Xf}(\nabla\phi\cdot\nabla X)^2+\frac{f_{,X}}{2Xf}\nabla^\mu\phi\nabla^\nu\phi\nabla_\mu\nabla_\nu X,\label{rho-vbhorn}\eea effective pressure:

\bea &&p_\text{eff}=\frac{f_{,\phi\phi}}{f}2X-\frac{2f_{,\phi}}{3f}\nabla^2\phi+\frac{1}{6f}\left(A-4f_{,XX}\right)(\nabla X)^2-\frac{2f_{,X}}{3f}\nabla^2X-\frac{1}{6Xf}\left(f_{,\phi}+12Xf_{,\phi X}\right)(\nabla\phi\cdot\nabla X)\nonumber\\
&&\;\;\;\;\;\;\;\;\;+\frac{1}{6Xf}\left(f_{,XX}-A\right)(\nabla\phi\cdot\nabla X)^2+\frac{f_{,X}}{6Xf}\nabla^\mu\phi\nabla^\nu\phi\nabla_\mu\nabla_\nu X,\label{p-vbhorn}\eea effective heat flux vector:

\bea &&q^\text{eff}_\mu=\frac{1}{\sqrt{2X}f}\left\{\left[f_{,\phi}+2Xf_{,\phi X}-(f_{,XX}-A)(\nabla\phi\cdot\nabla X)\right]\nabla_\mu-f_{,X}\nabla^\lambda\phi\nabla_\lambda\nabla_\mu\right\}X\nonumber\\
&&\;\;\;\;\;\;\;\;\;+\frac{1}{(2X)^{3/2}f}\left\{\left(f_{,\phi}+2Xf_{,\phi X}\right)(\nabla\phi\cdot\nabla X)-(f_{,XX}-A)(\nabla\phi\cdot\nabla X)^2-f_{,X}\nabla^\lambda\phi\nabla^\kappa\phi\nabla_\lambda\nabla_\kappa X\right\}\nabla_\mu\phi,\label{vbhorn-q}\eea and effective anisotropic stress tensor:

\bea \pi^\text{eff}_{\mu\nu}=\Pi^\text{eff}_{\mu\nu}-p_\text{eff}h_{\mu\nu},\label{vbhorn-pi}\eea where $p_\text{eff}$ is given by \eqref{p-vbhorn} and

\bea &&\Pi^\text{eff}_{\mu\nu}=\left[\frac{2f_{,\phi\phi}}{f}X-\frac{f_{,\phi}}{f}\nabla^2\phi-\frac{2f_{,\phi X}}{f}(\nabla\phi\cdot\nabla X)-\frac{2f_{,XX}-A}{2f}(\nabla X)^2-\frac{f_{,X}}{f}\nabla^2X\right]h_{\mu\nu}\nonumber\\
&&\;\;\;\;\;\;\;\;+\frac{f_{,\phi}}{f}\left[\nabla_\mu\nabla_\nu\phi-\frac{\nabla_{(\mu}\phi\nabla_{\nu)}X}{X}-\frac{(\nabla\phi\cdot\nabla X)}{4X^2}\nabla_\mu\phi\nabla_\nu\phi\right]+\frac{f_{,XX}-A}{f}\left[\nabla_\mu X\nabla_\nu X+\frac{(\nabla\phi\cdot\nabla X)}{X}\nabla_{(\mu}\phi\nabla_{\nu)}X\right.\nonumber\\
&&\left.\;\;\;\;\;\;\;\;+\frac{(\nabla\phi\cdot\nabla X)^2}{4X^2}\nabla_\mu\phi\nabla_\nu\phi\right]+\frac{f_{,X}}{f}\left[\nabla_\mu\nabla_\nu X+\frac{\nabla^\lambda\phi\nabla_{(\mu}\phi\nabla_{\nu)}\nabla_\lambda X}{X}+\frac{\nabla^\lambda\phi\nabla^\kappa\phi\nabla_\lambda\nabla_\kappa X}{4X^2}\nabla_\mu\phi\nabla_\nu\phi\right].\label{vbhorn-Pi}\eea


\section{horndeski/beyond horndeski cosmological imperfect/perfect fluids}\label{sect-cosmo}

In sections \ref{sect-imp-fluid} and \ref{sect-via-bhorn} we have shown that, in the general case, both the viable Horndeski and beyond Horndeski theories admit an imperfect fluid representation. However, this is a correct statement only if consider a scalar field with non-vanishing spatial gradient. This means that the 4-velocity $u_\mu=\nabla_\mu\phi/\sqrt{-(\nabla\phi)^2}$ can not be that of free-falling observers. In a Friedmann-Robertson-Walker (FRW) cosmological framework, on the contrary, the timelike 4-velocity $u_\mu=\nabla_\mu\phi/\dot\phi=\delta^0_\mu$, is that of a free-falling observer. Hence, the acceleration of the co-moving observers vanishes since along geodesics, necessarily: $\dot u_\mu=0$. This is also true for the Horndeski and beyond Horndeski theories since the 4-velocity vector is the same: $u_\mu=\delta^0_\mu$. This is why the heat-flux vectors for the BD theory \eqref{q-flux}:

\bea q^\text{BD}_\mu=\frac{1}{\phi\sqrt{2X}}\left[\nabla_\mu X+\frac{(\nabla\phi\cdot\nabla X)}{2X}\nabla_\mu\phi\right]=-\frac{1}{\phi}\sqrt{2X}\dot u_\mu,\label{q-bd-discu}\eea and for the viable Horndeski theories:

\bea q^\text{vhorn}_\mu=\frac{G_{4,\phi}-XG_{3,X}}{\sqrt{2X}G_4}\left[\nabla_\mu X+\frac{(\nabla\phi\cdot\nabla X)}{2X}\nabla_\mu\phi\right]=-\frac{G_{4,\phi}-XG_{3,X}}{G_4}\sqrt{2X}\dot u_\mu,\label{q-vhorn-discu}\eea both vanish: $q^\text{BD}_\mu=q^\text{vhorn}_\mu=0$. Above we have taken into account the expression \eqref{accel-x} for the acceleration:

\bea \dot u_\mu=u^\lambda\nabla_\lambda u_\mu=-\frac{1}{2X}\left[\nabla_\mu X+\frac{(\nabla\phi\cdot\nabla X)}{2X}\nabla_\mu\phi\right].\nonumber\eea

For the beyond Horndeski theories we have that the heat flux vector \eqref{vbhorn-q} can be written as:

\bea &&q^\text{vbhorn}_\mu=\frac{f_{,\phi}+4Xf_{,\phi X}-\left(f_{,XX}-A\right)(\nabla\phi\cdot\nabla X)}{\sqrt{2X}f}\left[\nabla_\mu X+\frac{(\nabla\phi\cdot\nabla X)}{2X}\nabla_\mu\phi\right]\nonumber\\
&&\;\;\;\;\;\;\;\;\;\;\;\;\;\;\;-\frac{f_{,X}}{\sqrt{2X}f}\nabla^\lambda\phi\left[\nabla_\lambda\nabla_\mu X+\frac{\nabla^\kappa\phi\nabla_\lambda\nabla_\kappa X}{2X}\nabla_\mu\phi\right],\nonumber\eea but, since $u^\mu=\nabla^\mu\phi/\sqrt{2X}=g^{0\mu}$, then the second term in the RHS of the above equation exactly vanishes:

\bea -\frac{f_{,X}}{f}g^{0\lambda}\left[\nabla_\lambda\nabla_\mu X+g^{0\kappa}\delta^0_\mu\nabla_\lambda\nabla_\kappa X\right]=-\frac{f_{,X}}{f}\left[\nabla^0\dot X+\ddot X\right]\delta^0_\mu=-\frac{f_{,X}}{f}\left[-\ddot X+\ddot X\right]\delta^0_\mu=0,\nonumber\eea so that

\bea q^\text{vbhorn}_\mu=-\frac{f_{,\phi}+4Xf_{,\phi X}-\left(f_{,XX}-A\right)(\nabla\phi\cdot\nabla X)}{f}\sqrt{2X}\dot u_\mu=0,\label{q-vbhorn-discu}\eea as it was for the BD and the viable Horndeski theories.

Let us assume a FRW spacetime with metric: $ds^2=-dt^2+a^2(t)\delta_{ij}dx^idx^j$ ($i,j=1,2,3$). As usual $t$ is the cosmic time, $a=a(t)$ is the scale factor and $H\equiv\dot a/a$ is the Hubble parameter. Here an over-dot accounts for the derivative in respect to the cosmic time. The timelike FRW 4-velocity is given by: $u_\mu=\nabla_\mu\phi/\dot\phi=\delta^0_\mu$, so that for the components of the 3-metric $h_{\mu\nu}$ we obtain: $h_{00}=0$, $h_{ij}=a^2(t)\delta_{ij}$. We also have:

\bea X=\dot\phi^2/2,\;\nabla^2\phi=-(\ddot\phi+3H\dot\phi),\;(\nabla\phi\cdot\nabla X)=-\dot\phi^2\ddot\phi,\;\nabla_\mu\nabla_\nu\phi=\ddot\phi\delta^0_\mu\delta^0_\nu-H\dot\phi h_{\mu\nu},\label{useful}\eea so that, by making the appropriate substitutions, we get the following expressions for the effective kinematic quantities of viable Horndeski theories \eqref{vhorn-lag} in FRW spacetimes:

\bea \rho_\text{eff}=\frac{1}{2G_4}\left[\left(K_{,X}-G_{3,\phi}+3G_{3,X}H\dot\phi\right)\dot\phi^2-K-6G_{4,\phi}H\dot\phi\right],\label{vhorn-rho}\eea

\bea p_\text{eff}=\frac{1}{2G_4}\left[\left(2G_{4,\phi\phi}-G_{3,\phi}-G_{3,X}\ddot\phi\right)\dot\phi^2+K+2G_{4,\phi}\left(\frac{5}{3}\ddot\phi+2H\dot\phi\right)\right],\label{vhorn-p}\eea

\bea \pi^\text{eff}_{\mu\nu}=-\frac{2G_{4,\phi}}{3G_4}\ddot\phi h_{\mu\nu},\label{vhorn-pimn}\eea and, as shown above: $q^\text{eff}_\mu=0$. Since $h_{00}=0$, the anisotropic stress has only spatial non-vanishing components.

It is noticeable that the higher-derivative terms (the cubic term in the present case) do not contribute neither towards the effective heat flux vector nor to the anisotropic stresses. Hence, in a cosmological setup, only the non-minimal coupling $G_4(\phi)$ of the scalar field to the curvature scalar makes of the effective SET \eqref{case-study} the one of an imperfect fluid like in \eqref{eff-set}.

In what regards to the viable beyond Horndeski effective cosmological fluid we have that, for the effective energy density:

\bea &&\rho_\text{eff}=\frac{3f_{,X}}{f}\left(\dot H+2H^2\right)\dot\phi^2-\frac{3f_{,\phi}}{f}H\dot\phi-\frac{3}{f}\left(f_{,X}-A\dot\phi^2\right)H\dot\phi\ddot\phi\nonumber\\
&&\;\;\;\;\;\;\;\;+\frac{1}{f}\left(2f_{,\phi X}+A_{,\phi}\dot\phi^2\right)\dot\phi^2\ddot\phi+\frac{1}{2f}\left(A+A_{,X}\dot\phi^2\right)\dot\phi^2\ddot\phi^2+\frac{A}{f}\dot\phi^3\dddot\phi,\label{rho-vbhorn-frw}\eea while for the effective pressure:

\bea &&p_\text{eff}=\frac{2f_{,\phi}}{f}H\dot\phi+\frac{2f_{,X}}{f}H\dot\phi\ddot\phi+\frac{f_{,\phi\phi}}{f}\dot\phi^2+\frac{1}{f}\left(f_{,\phi}+2\dot\phi^2f_{,\phi X}\right)\ddot\phi\nonumber\\
&&\;\;\;\;\;\;\;\;+\frac{1}{2f}\left(2f_{,XX}-A\right)\dot\phi^2\ddot\phi^2+\frac{f_{,X}}{f}\left[\ddot\phi^2+\dot\phi(\dddot\phi)\right],\label{p-vbhorn-frw}\eea where we have taken into account that for the FRW metric the curvature scalar $R=6\left(\dot H+2H^2\right)$, while: 

\bea &&(\nabla X)^2=-\dot X^2=-(\dot\phi\ddot\phi)^2,\;\nabla_\mu\nabla_\nu X=\ddot X\delta^0_\mu\delta^0_\nu-H\dot Xh_{\mu\nu},\nonumber\\
&&\nabla^2X=-\ddot X-3H\dot X=-\ddot\phi(\ddot\phi+3H\dot\phi)-\dot\phi(\dddot\phi).\nonumber\eea For the flux vector, as shown at the beginning of this section, we get: $q^\text{eff}_\mu=0$, while since: 

\bea &&\Pi^\text{eff}_{\mu\nu}=\left\{\frac{2f_{,\phi}}{f}H\dot\phi+\frac{2f_{,X}}{f}H\dot\phi\ddot\phi+\frac{f_{,\phi\phi}}{f}\dot\phi^2+\frac{1}{f}\left(f_{,\phi}+2\dot\phi^2f_{,\phi X}\right)\ddot\phi\right.\nonumber\\
&&\left.\;\;\;\;\;\;\;\;\;\;+\frac{1}{2f}\left(2f_{,XX}-A\right)\dot\phi^2\ddot\phi^2+\frac{f_{,X}}{f}\left[\ddot\phi^2+\dot\phi(\dddot\phi)\right]\right\}h_{\mu\nu}=p_\text{eff}h_{\mu\nu},\label{Pi-vbhorn-frw}\eea the anisotropic stresses vanish as well: $\pi^\text{eff}_{\mu\nu}=0$. This means that in a FRW spacetime the viable beyond Horndeski theories are formally equivalent to GR with an effective perfect fluid. In this case the higher-order derivatives of the scalar field contribute only towards the effective energy density and the effective pressure of the perfect fluid, respectively.

Notice that in the cosmological framework the non-minimal coupling to the curvature scalar through the function $f=f(\phi,X)$ in \eqref{vbhorn-lag} is not enough to warrant the effective imperfect fluid representation. This is in contrast to the Horndeski theories \eqref{vhorn-lag}, where the non-minimal coupling through the function $G_4=G_4(\phi)$ is what allows to write the motion equations in the form of the GR Einstein's equations with an imperfect fluid. This means that, somehow, the higher-derivative terms with coupling function $A=A(\phi,X)$ in \eqref{vbhorn-lag} balance the contribution to the anisotropic stresses of the terms coming from the non-minimal coupling $f(\phi,X)R$ to the curvature scalar.


\section{discussion and conclusion}\label{sect-discuss}

In this paper we have demonstrated that a formal equivalence between viable Horndeski and beyond Horndeski theories and GR with an imperfect fluid can be established. Our results represent further generalization of previously published works on the effective fluid equivalences \cite{pimentel-cqg-1989, faraoni-prd-2018, adiez-ijmpd-2005, arroja-prd-2010, akhoury-jhep-2009, pujolas-jcap-2010, pujolas-jhep-2011, capozz-ijgmmp-2018, capozz-ijgmmp-2019, capozz-arxiv-2019}.

The effective fluid description of Horndeski and beyond Horndeski theories represents an alternative opportunity to deal with cosmological perturbations within the higher-derivative generalizations of scalar-tensor theories among others. Although we have demonstrated the mentioned imperfect/perfect fluid equivalence exclusively for the mentioned theories, the present results may be applied to other modifications of gravity such as the extended theories of gravity that are based in the Lagrangian ${\cal L}\propto F\left(R,\nabla^2 R,\nabla^4R,\ldots\nabla^{2k}R\right)$. As mentioned in the introduction, the ETG-s are equivalent to multi-STT \cite{wands-cqg-1994}. One example is the sixth-order gravity given by the choice: $F=R+\alpha R\nabla^2 R$, which is equivalent to Brans-Dicke theory with vanishing coupling parameter $\omega=0$, with a BD scalar field $\phi$ and an additional canonical scalar field $\vphi$ as matter source. Hence it can be put into the form of GR with a mixture of an effective imperfect and a perfect fluids. Another example is given by the $f(R)$ theory. This is a particular case of Horndeski theory \eqref{horn-action} when $K=-Rf_{,R}-f$, $G_4=f_{,R}$, $G_3=G_5=0$. Under the replacement $\phi\rightarrow f_{,R}$, $V(\phi)\rightarrow f_{,R}-f$, the $f(R)$ theory represents BD theory with vanishing coupling parameter ($\omega=0$). Hence, the above mentioned modifications of general relativity can be given the form of GR with an effective perfect/imperfect fluid. In particular, for the $f(R)$-theory the kinematic quantities that appear in the effective SET \eqref{set-imp-f} are those given by equations \eqref{set-pi-phi}-\eqref{q-flux} with the substitutions: $\omega=0$ and $\phi=f_{,R}$. If take into account that $\nabla_\mu\phi=f_{,RR}\nabla_\mu R$ $\rightarrow\;(\nabla\phi)^2=f^2_{,RR}(\nabla R)^2$, etc., we get the same expressions of Ref. \cite{faraoni-prd-2018}. Notice, in particular, that if make these substitutions in \eqref{4-vel}, we obtain the definition of the $4$-velocity in \cite{faraoni-prd-2018}: $u_\mu=\nabla_\mu R/\sqrt{-(\nabla R)^2}$. The effective fluid approach of $f(R)$ theories has been investigated also in Ref. \cite{arjona(fdr)-prd-2019} from the perspective of the cosmological perturbations.


An important aspect of the higher-derivative theories is related with the speed of propagation of scalar and tensor cosmological perturbations. For the Horndeski theories the speed of propagation of the gravitational waves is given by \cite{kobayashi-rpp-rev, defelice}:

\bea c^2_\text{gw}=\frac{G_4-X\left(\ddot\phi G_{5,X}+G_{5,\phi}\right)}{G_4-2XG_{4,X}-X\left(\dot\phi HG_{5,X}-G_{5,\phi}\right)}.\label{c2w}\eea Hence, after the simultaneous detection of gravitational waves GW170817 and the $\gamma$-ray burst GRB 170817A \cite{gw-grb}, leading to inferring that the speed of propagation of the tensor perturbations coincides with the speed of light in vacuum $c^2_\text{gw}=1$ (recall that in this paper we work in the units system where the speed of light in vacuum $c=1$), only the theories with $G_5=0$, $G_4=G_4(\phi)$ survive the observational checks. These are known as viable Horndeski theories \cite{kobayashi-rpp-rev}. In a similar fashion the only beyond Horndeski theories that survive the observational checks related to the GW170817 event are the viable beyond Horndeski theories given by the Lagrangian \eqref{vbhorn-lag}. Yet, the surviving higher-derivative generalizations of STT-s have to be consistent with the limits $0\leq c^2_s\leq 1$ on the squared sound speed (also the squared speed of propagation of the scalar perturbations): $c^2_s=p_{,X}/\rho_{,X}$, in order to avoid gradient instability and to obey causality \cite{quiros-ijmpd-rev, roy-ellis-grg-2007}. These additional bounds establish conditions on the derivatives.

For the viable Horndeski theories \eqref{vhorn-lag} we have that the effective pressure ($p^\text{eff}=p_{(2)}+p_{(3)}+p_{(4)}$) is given by the following expression:

\bea p^\text{eff}=\frac{1}{G_4}\left[\frac{K}{2}+\left(2G_{4,\phi\phi}-G_{3,\phi}\right)X+\left(G_{3,X}X-G_{4,\phi}\right)\frac{(\nabla\phi\cdot\nabla X)}{2X}-\frac{2}{3}G_{4,\phi}\nabla^2\phi\right],\label{p-eff-vhorn}\eea while for the effective energy density ($\rho^\text{eff}=\rho_{(2)}+\rho_{(3)}+\rho_{(4)}$):

\bea \rho^\text{eff}=\frac{1}{G_4}\left[XK_{,X}-\frac{K}{2}-G_{3,\phi}X+\left(G_{3,X}X-G_{4,\phi}\right)\frac{(\nabla\phi\cdot\nabla X)}{2X}+\left(G_{4,\phi}-G_{3,X}X\right)\nabla^2\phi\right].\label{rho-eff-vhorn}\eea Hence, the speed of propagation of the scalar perturbations in the viable Horndeski theories reads:

\bea c^2_s=\frac{p^\text{eff}_{,X}}{\rho^\text{eff}_{,X}}=\frac{\frac{K_{,X}}{2}-G_{3,\phi X}X-G_{3,\phi}+\left(G_{3,XX}X^2+G_{4,\phi}\right)\frac{(\nabla\phi\cdot\nabla X)}{2X^2}+2G_{4,\phi\phi}}{\frac{K_{,X}}{2}-G_{3,\phi X}X-G_{3,\phi}+\left(G_{3,XX}X^2+G_{4,\phi}\right)\frac{(\nabla\phi\cdot\nabla X)}{2X^2}+K_{,XX}X-(G_{3,X}+G_{3,XX}X)\nabla^2\phi}.\label{c2s-bhorn}\eea Causality ($c^2_s\leq 1$) leads to the following condition on the derivatives of the scalar field:

\bea \nabla^2\phi\leq\frac{XK_{,XX}-2G_{4,\phi\phi}}{XG_{3,XX}+G_{3,X}},\label{c-1}\eea while the absence of gradient instability ($c^2_s\geq 0$) requires that:

\bea (\nabla\phi\cdot\nabla X)\geq\frac{\left[2XG_{3,\phi X}+2G_{3,\phi}-K_{,X}-4G_{4,\phi\phi}\right]X^2}{X^2G_{3,XX}+G_{4,\phi}},\label{c-2}\eea and that \eqref{c-1} is satisfied. 

Similar conditions on the derivatives can be found for the viable beyond Horndeski theories. This means that these ``viable'' theories as a matter of fact can be non-viable if the above conditions on the squared sound speed are not satisfied. In other words: the bounds \eqref{c-1} and \eqref{c-2} amount to further constraints on the physical viability of Horndeski and beyond Horndeski theories, that already satisfy $c^2_\text{gw}=1$.  

Before concluding this paper we want to make a comment on imperfect viscous fluids. It is usually understood that the stress-energy tensor of a perfect fluid is written in the customary form:

\bea T^\text{(pf)}_{\mu\nu}=\left(\rho+p\right)u_\mu u_\nu+pg_{\mu\nu},\label{set-pf}\eea while that of an imperfect fluid should look like \eqref{set-imp-f}:

\bea T_{\mu\nu}^\text{(if)}=(\rho+p)u_\mu u_\nu+pg_{\mu\nu}+2q_{(\mu}u_{\nu)}+\pi_{\mu\nu}.\nonumber\eea However, one should be very careful with such classification of physical fluids. It is known that there are dissipative processes that generate entropy such as, for instance, the re-establishment of thermal equilibrium through particle decay or scattering of particles off each other or on another medium. In such cases the generation of entropy\footnote{In addition to the bounds considered above, the local generation of entropy by imperfect fluids represents another viability criterion to take into account: the generation of entropy should be positive in order to satisfy the 2nd law of thermodynamics.} can be related to the expansion rate or the local fluid velocity through a bulk viscosity term $\propto\nabla_\lambda u^\lambda$ \cite{visc}. The energy-momentum tensor of the bulk viscous fluid $T^*_{\mu\nu}$ is that of an imperfect fluid with a first-order deviation from the thermodynamic equilibrium \cite{mtw}:

\bea T^*_{\mu\nu}=\left(\rho+p^*\right)u_\mu u_\nu+p^*g_{\mu\nu},\label{bv-set}\eea where the whole pressure of the fluid $p^*=p+p_\text{bv}$, with $p_\text{bv}=-\xi\nabla_\lambda u^\lambda$ representing the viscous contribution to the pressure of the fluid ($\xi$ is the bulk viscous coefficient). But notice that, formally, this SET resembles the one of a perfect fluid \eqref{set-pf}. This is motivated by the fact that the imperfect fluid SET \eqref{set-imp-f} is contributed by heat fluxes and anisotropic pressure but it does not takes into account the contribution due to bulk viscosity. In this regard it could be interesting to check whether some of the effective perfect fluid SET-s already studied in the bibliography, in the context of extended theories of gravity \cite{capozz-ijgmmp-2018, capozz-ijgmmp-2019, capozz-arxiv-2019}, for instance, can accommodate first-order deviation from thermodynamic equilibrium leading to bulk viscosity, so that these can be traded as effective imperfect fluids.


\acknowledgments

Useful comments by Evgeny Novikov and Vasilis Oikonomou are acknowledged. The authors thank SNI-CONACyT for continuous support of their research activity. Ulises Nucamendi also acknowledges PRODEP-SEP and CIC-UMSNH for financial support of his contribution to the present research.


\begin{thebibliography}{99}



\bibitem{jordan-stt} P. Jordan, Astr. Nachr. {\bf 276} (1948) 193–208; P. Jordan, Z. Phys. {\bf 157} (1959) 112-121

\bibitem{bd-1961} C. Brans, R.H. Dicke, Phys. Rev. {\bf 124} (1961) 925-935

\bibitem{brans-phd-thesis} C.H. Brans, ''Mach's principle and a varying gravitational constant'', {\it Princeton dissertation}, May 1961

\bibitem{goenner-bd-history} H. Goenner, Gen. Rel. Grav. {\bf 44} (2012) 2077-2097 [arXiv:1204.3455]

 
\bibitem{fujii-book} Y. Fujii, K. Maeda, {\it The Scalar-Tensor Theory of Gravitation} (Cambridge University Press, 2003)

\bibitem{faraoni-book} V. Faraoni, {\it Cosmology in Scalar-Tensor Gravity} (Kluwer Academic Publishers, The Netherlands, 2004)

\bibitem{nordvedt-stt} K. Nordtvedt, Astrophys. J. {\bf 161} (1970) 1059-1067

\bibitem{wagoner-stt} R.V. Wagoner, Phys. Rev. D {\bf 1} (1970) 3209-3216

\bibitem{bergmann-stt} P.G. Bergmann, Int. J. Theor. Phys. {\bf 1} (1968) 25-36

\bibitem{reasenberg-stt} R.D. Reasenberg, I.I. Shapiro, P.E. MacNeil, R.B. Goldstein, J.C. Breidenthal, J.P. Brenkle, D.L. Cain, T.M. Kaufman, T.A. Komarek, A.I. Zygielbaum, Astrophys. J. {\bf 234} (1979) L219-L221

\bibitem{damour-stt} T. Damour, K. Nordtvedt, Phys. Rev. Lett. {\bf 70} (1993) 2217-2219; T. Damour, K. Nordtvedt, Phys. Rev. D {\bf 48} (1993) 3436-3450; T. Damour, G. Esposito-Farese, Phys. Rev. D {\bf 54} (1996) 1474-1491 [gr-qc/9602056]; T. Damour, Eur. Phys. J. C {\bf 3} (1998) 113-116

\bibitem{farese-polarski-prd-2001} G. Esposito-Farese, D. Polarski, Phys. Rev. D {\bf 63} (2001) 063504 [gr-qc/0009034]

\bibitem{anderson-yunes-prd-2017} D. Anderson, N. Yunes, Phys. Rev. D {\bf 96} (2017) 064037 [arXiv:1705.06351]


\bibitem{capozz-fdr} S. Capozziello, Int. J. Mod. Phys. D {\bf 11} (2002) 483-492 [gr-qc/0201033]; S. Capozziello, S. Carloni, A. Troisi, Recent Res. Dev. Astron. Astrophys. {\bf 1} (2003) 625 [astro-ph/0303041]

\bibitem{vollick-prd-2003} D.N. Vollick, Phys. Rev. D {\bf 68} (2003) 063510 [astro-ph/030663]

\bibitem{chiba-fdr} T. Chiba, Phys. Lett. B {\bf 575} (2003) 1-3 [astro-ph/0307338]; T. Chiba, T.L. Smith, A.L. Erickcek, Phys. Rev. D {\bf 75} (2007) 124014 [astro-ph/0611867]

\bibitem{carroll-fdr} S.M. Carroll, V. Duvvuri, M. Trodden, M.S. Turner, Phys. Rev. D {\bf 70} (2004) 043528 [astro-ph/0306438]; S.M. Carroll, A. De Felice, V. Duvvuri, D.A. Easson, M. Trodden, M.S. Turner, Phys. Rev. D {\bf 71} (2005) 063513 [astro-ph/0410031] 

\bibitem{sotiriou-cqg-2006} T.P. Sotiriou, Class. Quant. Grav. {\bf 23} (2006) 5117-5128 [gr-qc/0604028]

\bibitem{nojiri-odintsov-fdr} S. Nojiri, S.D. Odintsov, Gen. Rel. Grav. {\bf 36} (2004) 1765-1780 [hep-th/0308176]; S. Nojiri, S.D. Odintsov, Int. J. Geom. Meth. Mod. Phys. {\bf 4} (2007) 115-146 [hep-th/060121]; S. Nojiri, S.D. Odintsov, Phys. Lett. B {\bf 657} (2007) 238-245 [arXiv:0707.194]; S. Nojiri, S.D. Odintsov, Phys. Rev. D {\bf 77} (2008) 026007 [arXiv:0710.173]

\bibitem{cognola-prd-2008} G. Cognola, E. Elizalde, S. Nojiri, S.D. Odintsov, L. Sebastiani, S. Zerbini, Phys. Rev. D {\bf 77} (2008) 046009 [arXiv:0712.4017]

\bibitem{shaw-prd-2008} P. Brax, C. van de Bruck, A.C. Davis, D.J. Shaw, Phys. Rev. D {\bf 78} (2008) 104021 [arXiv:0806.3415]

\bibitem{sotiriou-rmp-2010} T.P. Sotiriou, V. Faraoni, Rev. Mod. Phys. {\bf 82} (2010) 451-497 [arXiv:0805.1726]

\bibitem{defelice-tsujikawa-lrr-2010} A. De Felice, S. Tsujikawa, Living Rev. Rel. {\bf 13} (2010) 3 [arXiv:1002.4928]

\bibitem{nojiri-odintsov-phys-rep-2011} S. Nojiri, S.D. Odintsov, Phys. Rept. {\bf 505} (2011) 59-144 [arXiv:1011.0544]

\bibitem{vasilis-phys-rep-2017} S. Nojiri, S.D. Odintsov, V.K. Oikonomou, Phys. Rept. {\bf 692} (2017) 1-104 [arXiv:1705.11098]


\bibitem{chiba-jcap-2005} T. Chiba, J. Cosmol. Astropart. Phys. {\bf 0503} (2005) 008 [arXiv:gr-qc/0502070]

\bibitem{capozziello-grg-rev} S. Capozziello, M. Francaviglia, Gen. Rel. Grav. {\bf 40} (2008) 357-420 [arXiv:0706.1146]

\bibitem{capoz-phys-rept-rev} S. Capozziello, M. De Laurentis, Phys. Rept. {\bf 509} (2011) 167-321 [arXiv:1108.6266]

\bibitem{gottlober-cqg-1990} S. Gottlober, H.J. Schmidt, A.A. Starobinsky, Class. Quant. Grav. {\bf 7} (1990) 893

\bibitem{schmidt-cqg-1990} H.J. Schmidt, Class. Quant. Grav. {\bf 7} (1990) 1023-1031

\bibitem{wands-cqg-1994} D. Wands, Class. Quant. Grav. {\bf 11} (1994) 269-280 [gr-qc/9307034]

\bibitem{capoz-etg-grg-2000} S. Capozziello, G. Lambiase, Gen. Rel. Grav. {\bf 32} (2000) 295-311 [gr-qc/9912084]

\bibitem{nojiri-ijgmp-2007} S. Nojiri, S.D. Odintsov, Int. J. Geom. Meth. Mod. Phys. {\bf 4} (2007) 115-146 [hep-th/0601213]

\bibitem{capoz-etg-prd-2015} S. Capozziello, F.S.N. Lobo, J.P. Mimoso, Phys. Rev. D {\bf 91} (2015) 124019 [arXiv:1407.7293]


\bibitem{horndeski} G.W. Horndeski, Int. J. Theor. Phys. {\bf 10} (1974) 363-384

\bibitem{nicolis-gal} A. Nicolis, R. Rattazzi, E. Trincherini, Phys. Rev. D {\bf 79} (2009) 064036 [arXiv:0811.2197]

\bibitem{deffayet-vikman-gal} C. Deffayet, G. Esposito-Farese, A. Vikman, Phys. Rev. D {\bf 79} (2009) 084003 [arXiv:0901.1314]

\bibitem{deffayet-deser-gal} C. Deffayet, S. Deser, G. Esposito-Farese, Phys. Rev. D {\bf 80} (2009) 064015 [arXiv:0906.1967]; C. Deffayet, X. Gao, D.A. Steer, G. Zahariade, Phys. Rev. D {\bf 84} (2011) 064039 [arXiv:1103.3260]

\bibitem{kazuya-silva-gal} F.P. Silva, K. Koyama, Phys. Rev. D {\bf 80} (2009) 121301 [arXiv:0909.4538]

\bibitem{fab-4-prl-2012} C. Charmousis, E.J. Copeland, A. Padilla, P.M. Saffin, Phys. Rev. Lett. {\bf 108} (2012) 051101 [arXiv:1106.2000]

\bibitem{clifton-phys-rept-rev} T. Clifton, P.G. Ferreira, A. Padilla, C. Skordis, Phys. Rept. {\bf 513} (2012) 1-189 [arXiv:1106.2476]

\bibitem{deffayet-rev} C. Deffayet, D.A. Steer, Class. Quant. Grav. {\bf 30} (2013) 214006 [arXiv:1307.2450]

\bibitem{tsujikawa-lect-not} S. Tsujikawa, Lect. Notes Phys. {\bf 892} (2015) 97-136 [arXiv:1404.2684]

\bibitem{kazuya-rpp-2016} K. Koyama, Rept. Prog. Phys. {\bf 79} (2016) 046902 [arXiv:1504.04623]

\bibitem{quiros-ijmpd-rev} I. Quiros, Int. J. Mod. Phys. D {\bf 28} (2019) 1930012 [arXiv:1901.08690]

\bibitem{kobayashi-rpp-rev} T. Kobayashi, Rept. Prog. Phys. {\bf 82} (2019) 086901 [arXiv:1901.07183]


\bibitem{bhorn-langlois} J. Gleyzes, D. Langlois, F. Piazza, F. Vernizzi, Phys. Rev. Lett. {\bf 114} (2015) 211101 [arXiv:1404.6495]; J. Gleyzes, D. Langlois, F. Piazza, F. Vernizzi, JCAP {\bf 1502} (2015) 018 [arXiv:1408.1952]; D. Langlois, K. Noui, JCAP 1602 (2016) 034 [arXiv:1510.06930];  D. Langlois, K. Noui, JCAP {\bf 1607} (2016) 016 [arXiv:1512.06820]

\bibitem{bhorn-fasiello} M. Fasiello, S. Renaux-Petel, JCAP {\bf 1410} (2014) 037 [arXiv:1407.7280]

\bibitem{bhorn-crisostomi} M. Crisostomi, M. Hull, K. Koyama, G. Tasinato, JCAP {\bf 1603} (2016) 038 [arXiv:1601.04658]; J.B. Achour, M. Crisostomi, K. Koyama, D. Langlois, K. Noui, G. Tasinato, JHEP {\bf 1612} (2016) 100 [arXiv:1608.08135]

\bibitem{bhorn(vainsh)} T. Kobayashi, Y. Watanabe, D. Yamauchi, Phys. Rev. D {\bf 91} (2015) 064013 [arXiv:1411.4130]

\bibitem{mancarella-jcap-2017} D. Langlois, M. Mancarella, K. Noui, F. Vernizzi, JCAP {\bf 1705} (2017) 033 [arXiv:1703.03797]

\bibitem{bhorn-ostrog} C. de Rham, G. Gabadadze, A.J. Tolley, JHEP {\bf 1111} (2011) 093 [arXiv:1108.4521]; M. Zumalac\'arregui, J. Garc\'ia-Bellido, Phys. Rev. D {\bf 89} (2014) 064046 [arXiv:1308.4685]

\bibitem{chagoya-tasinato-jhep-2017} J. Chagoya, G. Tasinato, JHEP {\bf 1702} (2017) 113 [arXiv:1610.07980]



\bibitem{bailyn-prd-1980} M. Bailyn, Phys. Rev. D {\bf 22} (1980) 267-279

\bibitem{madsen-ass-1985} M.S. Madsen, Astrophys. Space Sci. 113 (1985) 205-207

\bibitem{faraoni-prd-2012} V. Faraoni, Phys. Rev. D {\bf 85} (2012) 024040 [arXiv:1201.1448]

\bibitem{semiz-prd-2012} I. Semiz, Phys. Rev. D {\bf 85} (2012) 068501
 
\bibitem{adiez-plb-2013} A. Diez-Tejedor, Phys. Lett. B {\bf 727} (2013) 27-30 [arXiv:1309.4756]

\bibitem{faraoni-ejpc-2019} V. Faraoni, Jeremy Cot\'e, Eur. Phys. J. C {\bf 79} (2019) 318 [arXiv:1812.06457]

\bibitem{madsen-cqg-1988} M.S. Madsen, Class. Quant. Grav. {\bf 5} (1988) 627-639

\bibitem{pimentel-cqg-1989} L.O. Pimentel, Class. Quant. Grav. {\bf 6} (1989) L263-L265

\bibitem{faraoni-prd-2018} V. Faraoni, J. Cot\'e, Phys. Rev. D {\bf 98} (2018) 084019 [arXiv:1808.02427]



\bibitem{adiez-ijmpd-2005} A. Diez-Tejedor, A. Feinstein, Int. J. Mod. Phys. D {\bf 14} (2005) 1561-1576 [gr-qc/0501101];

\bibitem{arroja-prd-2010} F. Arroja, M. Sasaki, Phys. Rev. D {\bf 81} (2010) 107301 [arXiv:1002.1376]

\bibitem{akhoury-jhep-2009} R. Akhoury, C.S. Gauthier, A. Vikman, JHEP {\bf 0903} (2009) 082 [arXiv:0811.1620]
 
\bibitem{pujolas-jcap-2010} C. Deffayet, O. Pujolas, I. Sawicki, A. Vikman, JCAP {\bf 1010} (2010) 026 [arXiv:1008.0048]

\bibitem{pujolas-jhep-2011} O. Pujolas, I. Sawicki, A. Vikman (CERN), JHEP {\bf 1111} (2011) 156 [arXiv:1103.5360]

\bibitem{capozz-ijgmmp-2018} S. Capozziello, C.A. Mantica, L.G. Molinari, Int. J. Geom. Meth. Mod. Phys. {\bf 16} (2018) 1950008 [arXiv:1810.03204]

\bibitem{capozz-ijgmmp-2019} S. Capozziello, C.A. Mantica, L.G. Molinari, Int. J. Geom. Meth. Mod. Phys. {\bf 16} (2019) 1950133 [arXiv:1906.05693]

\bibitem{capozz-arxiv-2019} S. Capozziello, C.A. Mantica, L.G. Molinari, arXiv:1908.10176

\bibitem{novikov} E.A. Novikov, Electron. J. Theor. Phys. {\bf 13} (2016) 79-90; Mod. Phys. Lett. A {\bf 31} (2016) 1650092

\bibitem{ellis} G.F.R. Ellis, Gen. Rel. Grav. {\bf 41} (2009) 581-660, Proc. Int. Sch. Phys. Fermi {\bf 47} (1971) 104-182

\bibitem{ray-jmp-1972} J.R. Ray, J. Math. Phys. {\bf 13} (1972) 1451-1453


\bibitem{gw-grb} B.P. Abbott et al (LIGO Scientific and Virgo Collaborations), Phys. Rev. Lett. {\bf 119} (2017) 161101 [arXiv:1710.05832]; B.P. Abbott et al (LIGO Scientific and Virgo and Fermi-GBM and INTEGRAL Collaborations), Astrophys. J. {\bf 848} (2017) L13 [arXiv:1710.05834] 

\bibitem{arjona-prd-2019} R. Arjona, W. Cardona, S. Nesseris, Phys. Rev. D {\bf 100} (2019) 063526 [arXiv:1904.06294]

\bibitem{deffayet_prd_2011} C. Deffayet, X. Gao, D.A. Steer, G. Zahariade, {\it Phys. Rev. D} {\bf 84} (2011) 064039 [arXiv:1103.3260]

\bibitem{tsujikawa_lect_not} S. Tsujikawa, Lect. Notes Phys. {\bf 892} (2015) 97-136 [arXiv:1404.2684]

\bibitem{kazuya_gal} F.P. Silva, K. Koyama, Phys. Rev. D {\bf 80} (2009) 121301 [arXiv:0909.4538]

\bibitem{cov-gal} C. Deffayet, G. Esposito-Farese, A. Vikman, Phys. Rev. D {\bf 79} (2009) 084003 [arXiv:0901.1314]; C. Deffayet, S. Deser, G. Esposito-Farese, Phys. Rev. D {\bf 80} (2009) 064015 [arXiv:0906.1967]; C. Deffayet, X. Gao, D.A. Steer, G. Zahariade, Phys. Rev. D {\bf 84} (2011) 064039 [arXiv:1103.3260]; C. Deffayet, D.A. Steer, Class. Quant. Grav. {\bf 30} (2013) 214006 [arXiv:1307.2450]

\bibitem{bhorn} M. Zumalac\'arregui, J. Garc\'ia-Bellido, Phys. Rev. D {\bf 89} (2014) 064046 [arXiv:1308.4685]; D. Langlois, K. Noui, JCAP {\bf 1602} (2016) 034 [arXiv:1510.06930]; JCAP {\bf 1607} (2016) 016 [arXiv:1512.06820]; J. Ben Achour, M. Crisostomi, K. Koyama, D. Langlois, K. Noui, G. Tasinato, JHEP {\bf 1612} (2016) 100 [arXiv:1608.08135]; M. Crisostomi, K. Koyama, G. Tasinato, JCAP 1604 (2016) 044 [arXiv:1602.03119]; M. Crisostomi, M. Hull, K. Koyama, G. Tasinato, JCAP {\bf 1603} (2016) 038 [arXiv:1601.04658]

\bibitem{arjona(fdr)-prd-2019} R. Arjona, W. Cardona, S. Nesseris, Phys. Rev. D {\bf 99} (2019) 043516 [arXiv:1811.02469]

\bibitem{defelice} A. De Felice, T. Kobayashi, S. Tsujikawa, Phys. Lett. B {\bf 706} (2011) 123-133 [arXiv:1108.4242]

\bibitem{roy-ellis-grg-2007} G. Ellis, R. Maartens, M.A.H. MacCallum, Gen. Rel. Grav. {\bf 39} (2007) 1651-1660 [gr-qc/0703121]

\bibitem{visc} J.R. Wilson, G.J. Mathews, G.M. Fuller, Phys. Rev. D {\bf 75} (2007) 043521 [astro-ph/0609687]

\bibitem{mtw} C.W. Misner, K.S. Thorne, J.A. Wheeler, {\it Gravitation}, (1973, W.H. Freeman and Company) pag. 567









\end{thebibliography}
\end{document}